\documentclass{optica-article}

\journal{opticajournal} 

\articletype{Research Article}

\usepackage{lineno}

\begin{document}

\title{High-quality amorphous Silicon Carbide for hybrid photonic integration at low temperature}

\author{Bruno Lopez-Rodriguez,\authormark{1,5,*} Roald van der Kolk,\authormark{2,5} Samarth Aggarwal\authormark{3,5}, Naresh Sharma\authormark{1}, Zizheng Li\authormark{1}, Daniel van der Plaats\authormark{2}, Thomas Scholte\authormark{1}, Jin Chang\authormark{4}, Silvania F. Pereira\authormark{1}, Simon Gr\"{o}blacher\authormark{4}, Harish Bhaskaran\authormark{3} and Iman Esmaeil Zadeh\authormark{1} }

\address{\authormark{1} Department of Imaging Physics (ImPhys), Faculty of Applied Sciences, Delft University of Technology, Delft 2628 CJ, The Netherlands\\
\authormark{2} Kavli Institute of Nanoscience, Delft University of Technology, Delft 2628 CD, The Netherlands\\
\authormark{3} Department of Materials, University of Oxford, Parks Road, Oxford OX1 3PH, U.K.\\
\authormark{4} Department of Quantum Nanoscience, Faculty of Applied Sciences, Delft University of Technology, Delft 2628 CJ, The Netherlands\\
\authormark{5} The authors contributed equally to this work\\}

\email{\authormark{*}b.lopezrodriguez@tudelft.nl} 


\begin{abstract*} 
Integrated photonic platforms have proliferated in recent years, each demonstrating its own unique strengths and shortcomings. However, given the processing incompatibilities of different platforms, a formidable challenge in the field of integrated photonics still remains for combining the strength of different optical materials in one hybrid integrated platform. Silicon carbide is a material of great interest because of its high refractive index, strong second and third-order non-linearities and broad transparecy window in the visible and near infrared. However, integrating SiC has been difficult, and current approaches rely on transfer bonding techniques, that are time consuming, expensive and lacking precision in layer thickness. Here, we demonstrate high index Amorphous Silicon Carbide (a-SiC) films deposited at 150$^{\circ}$C and verify the high performance of the platform by fabricating standard photonic waveguides and ring resonators. The intrinsic quality factors of single-mode ring resonators were in the range of $Q_{int} = (4.7-5.7)\times10^5$ corresponding to optical losses between 0.78-1.06 dB/cm. We then demonstrate the potential of this platform for future heterogeneous integration with ultralow loss thin SiN and LiNbO$_3$ platforms. 

\end{abstract*}

\section{Introduction}

Integrated photonics is a rapidly-growing field that is revolutionizing the way we use light for computing, communication, and sensing. By developing new platforms and technologies, researchers are continuously enhancing the performance and capabilities of the building blocks of future photonic technologies. Silicon-On-Insulator\cite{silicon-photonics}, Silicon Nitride\cite{Xiang:22} and Aluminum Nitride\cite{aluminium-nitride} have shown outstanding performance, for example, sub-picometer wavelength filters, low loss and high visibility Mach-Zehnder interferometers and accurate variable beam splitters. Due to the persisting demand to unlock new properties and allow for higher degrees of freedom in photonic devices, materials that offer tunability and strong non-linear behavior have gained attention in recent years. 

Silicon Carbide (SiC) is emerging as a promising material for integrated quantum photonics due to its unique characteristics such as a high refractive index, strong second- and third-order optical non-linearities \cite{Sato:09,Zheng:19} and a broad transparency window from visible to the mid-infrared range \cite{midir-sic}. For quantum computing experiments, different crystalline forms of silicon carbide are being incorporated in a broad range of photonic schemes to individually address single-photon sources\cite{Lukin_2020} and spin-qubits \cite{sic-qubits}. 4H-SOI SiC ring resonators have been shown to exhibit quality factors of up to $1.1\times10^6$, making them a valuable demonstrator for optical parametric oscillation \cite{Guidry:20}.  On the other hand, the highest reported quality factor in a silicon carbide platform was achieved using its crystalline form 4H-SiCOI and reached values up to $6.75\times10^6$. However, one challenge in using SiC in quantum photonics is the need for transfer-bonding methods when depositing the crystalline material onto other substrates \cite{Wang2021} involving expensive and time-consuming processes together with issues regarding precise thickness control, complicating hybrid integration. Furthermore, provided that this last requirement is fulfilled, processing temperatures and chemical interactions between the different materials give rise to compatibility issues.      

For hybrid integration with other platforms, inert material such as amorphous silicon carbide has great potential. One of the most promising properties of a-SiC is its strong third-order nonlinearity, which is ten times higher than SiN \cite{sin-kerr} and crystalline SiC \cite{csic-kerr}, useful in, for example, four-wave mixing processes. From the standard Chemical Vapor Deposition (CVD) techniques, Plasma-Enhanced CVD (PECVD) have shown excellence in terms of optical performances in ring resonators with intrinsic quality factor reaching up to $1.6\times 10^5$ at around 1550 nm\cite{cmos-asic}. Furthermore, it was shown compatibility with the well-established CMOS fabrication processes. Recently, four-wave mixing has also been demonstrated using this platform with micro-ring resonators having loaded quality factors of $0.7\times10^5$ at around 1550 nm \cite{fwm-asic}. Therefore, it remains a challenge to decrease the losses in this platform and compete with well-established technologies.

In this work, we fabricate and characterize ring resonators on amorphous silicon carbide films deposited via Inductively Coupled Plasma-Enhanced CVD (Oxford ICPCVD PlasmaPro100). All optical devices show intrinsic quality factors above $4.66\times 10^5$, with the highest being $5.7\times 10^5$, overall, more than three times higher than previous achievements with this material and waveguide propagation loss ranging between 0.78-1.07 dB/cm. Additionally, using our ICPCVD optimized recipe, the a-SiC films can be deposited at 150$^{\circ}$C, which to our knowledge is the lowest temperature among other techniques and can be implemented with a variety of optical materials with a simple lift-off process. Most importantly, we demonstrate a fabrication route for heterogeneous integration of a-SiC films with SiN and LNOI supported by optical simulations. 

\subsection{\label{sec:level2}Amorphous Silicon Carbide}

\subsubsection{\label{sec:level3}Deposition of a-SiC films}

Amorphous silicon carbide has gained interest as a photonic platform due to its high refractive index, large and tunable band gap, chemically inert nature, and potential compatibility with CMOS processes. The deposition of amorphous Silicon Carbide thin films can be achieved with Low-Pressure CVD (LPCVD) \cite{MORANA2013654}, PECVD \cite{CHEN20149791} and ICPCVD \cite{FRISCHMUTH2016647}, where the latest two have been shown in previous studies with good reproducibility not only for Silicon Carbide but also with other materials for photonic devices, such as Silicon Nitride. \cite{Dawn2019,Wu:18,Ji:19}.

The main difference between PECVD and ICPCVD is in the plasma coupling mechanisms, i.e. inductive coupling in the case of ICPCVD while PECVD is capacitively coupled. In the case of PECVD, the bias between the parallel plates is coupled to the forward plasma power, which in turn means that a higher plasma density can cause more ion damage to the substrate compared to ICPCVD. Therefore, in PECVD, the plasma densities have to be kept lower than in ICPCVD. The latter means that in ICPCVD depositions, lower deposition temperatures and higher densities can be achieved\cite{doi:https://doi.org/10.1002/0471724254.ch6, doi:https://doi.org/10.1002/0471724254.ch11, doi:https://doi.org/10.1002/0471724254.ch12}.

\begin{figure}[ht!]
\includegraphics[width=1\textwidth]{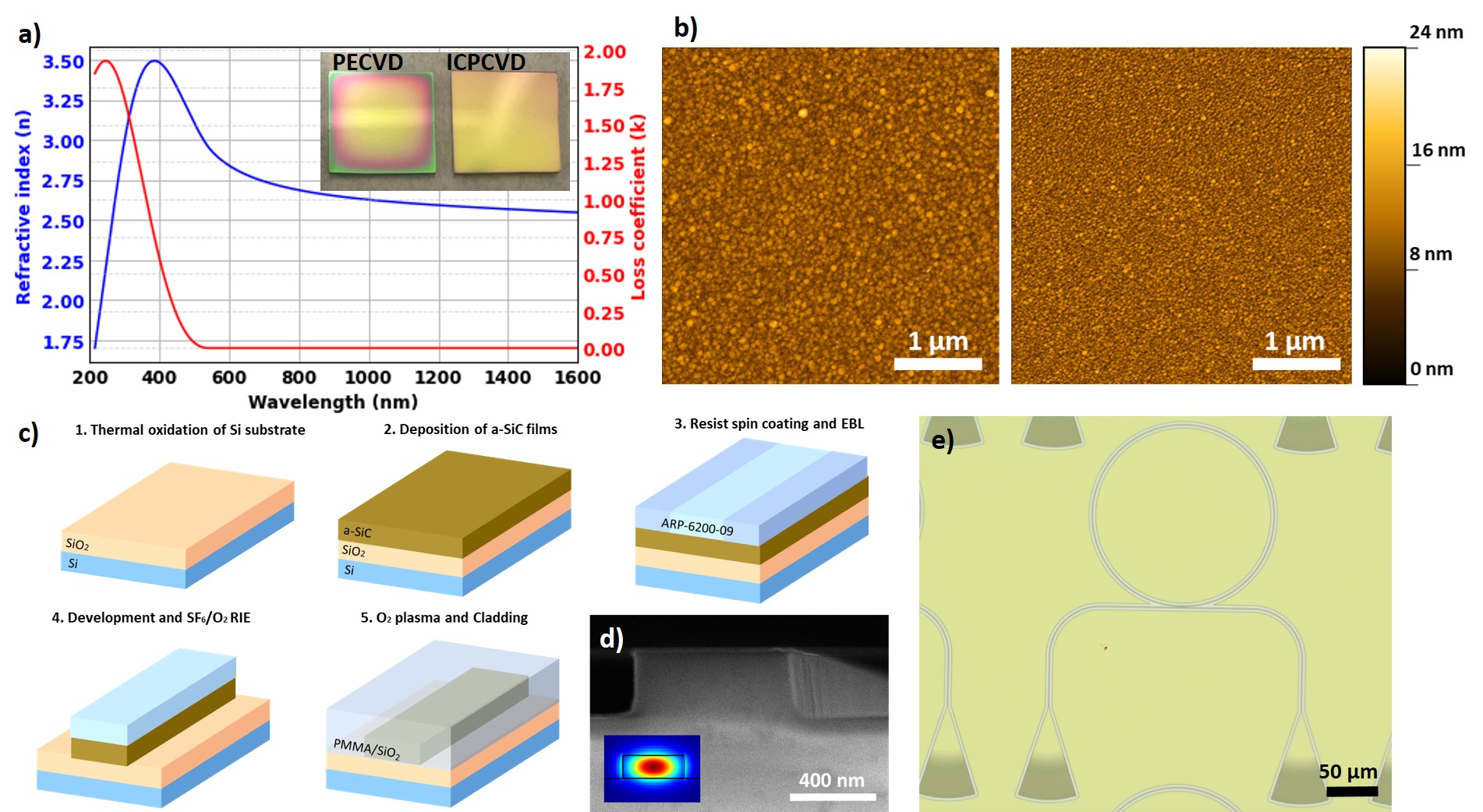}

\caption{\label{fig:afm-elip} a) Refractive index (n) and loss coefficient (k) for ICPCVD films deposited at 150$^{\circ}$C. Inset: comparison between film uniformities for the cases of deposition using PECVD and ICPCVD techniques, b) AFM scans of films deposited at 250$^{\circ}$C using PECVD (left) and ICPCVD (right), c) fabrication flow for the optical devices and d) SEM image of the waveguide cross-section. Inset: FDTD simulation of the confined mode and e) optical microscope image of a ring resonator device with grating couplers.} 

\end{figure}

In optical devices, the performance is mainly affected by the presence of Si-H and N-H bonds, which is the major loss mechanism for SiN-based resonators (assuming that roughness effects have been eliminated through conventional techniques) \cite{Pfeiffer:18,Jin_2021}. In the case of a-SiC:H thin films, Si-H bonds are also present in addition to C-H bonds. As shown in some studies, hydrogenation decreases with density, involving the increase of the deposition temperature \cite{Karouta2012}. 


Fig.\ref{fig:afm-elip}a shows the experimentally measured (using an ellipsometer) refractive index (n) and the loss coefficient (k) of an a-SiC film deposited at 150$^{\circ}$C. The refractive index of the a-SiC is 2.55 at 1550 nm, which is higher than the refractive index of SiN, leading to more compact and improved device integration since a higher refractive index translates into a high field confinement. Most importantly, depending on the Si and C content of the films, the refractive index and the overall properties of the material can be tuned to match the specific requirements (see Fig. S1 in the supplementary information).  

The inset of fig.\ref{fig:afm-elip}a shows typical deposition results on a small 15x15 mm thermally oxidized silicon sample. The color variation close to the edges reveals thickness non-uniformity in the PECVD sample (left) due to thin film interference \cite{Wang_Li_2013} while, in contrast, ICPCVD (right) shows excellent uniformity. The non-uniformity in PECVD is primarily attributed to edge effects and skin effects \cite{Chabert_2007}, which are more prominent in smaller samples due to their increased surface-to-volume ratio. Moreover, the larger plasma sheet leads to more ions accelerating towards the sample from the edge regions and with higher energies\cite{doi:https://doi.org/10.1002/0471724254.ch6}. Such uniformity is especially important for wafer-scale processing, where the cost can be reduced through the optimization of the deposition process. 

Atomic Force Microscopy (AFM) images in fig.\ref{fig:afm-elip}b reveal that the grain size of the PECVD film (left) is significantly larger than that of the ICPCVD film (right). This observation is consistent with previous studies that have suggested that higher plasma densities in PECVD lead to larger grain sizes\cite{COSCIA2005433}. Specifically, the root-mean-square (r$_q$) values for the ICPCVD and PECVD films were found to be 1.02 nm and 1.27 nm, respectively. The difference in surface roughness and grain size between the two films can significantly affect their optical properties.

An important advantage of a-SiC is the possibility to incorporate nitrogen as has been previously demonstrated\cite{SAFRANKOVAA1998165} that could lead to conductive films and optical elements where the devices can be tuned directly with electrical contacts, and therefore allowing configurations for e.g. optical switches\cite{LI20105} or adding tunability to multimode interferometers (MMIs)\cite{mmi}.

\subsection{\label{sec:level4}EXPERIMENTAL METHODS}

\begin{figure*}
\includegraphics[width=1\textwidth]{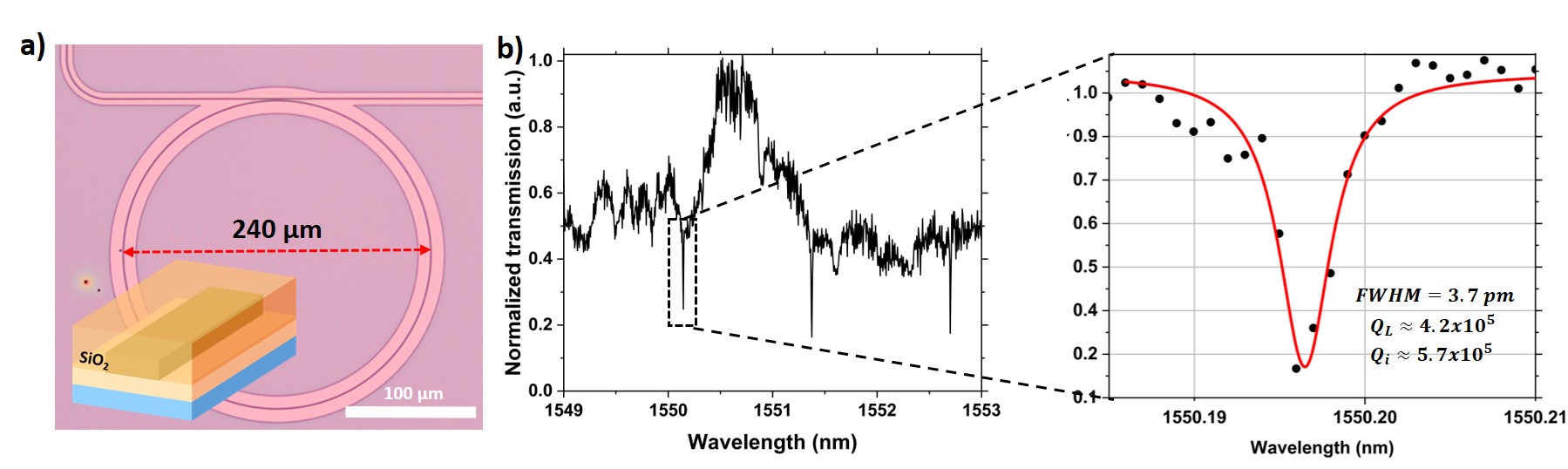}
\caption{\label{fig:best_q}a) Optical microscope image of a ring resonator made on films deposited at 150$^{\circ}$C with ICPCVD covered with silicon dioxide cladding as represented in the inset, b) spectrum between 1549 nm and 1553 nm of the device and scan with 1 pm resolution of the selected resonant dip at 1550.196 nm}

\end{figure*}

A complete process flow of the fabrication of the final photonic devices is shown in fig.\ref{fig:afm-elip}c. a-SiC films were deposited with ICPCVD on 2.5 $\mu$m thermally grown silicon dioxide. The film thickness for the deposited a-SiC was chosen according to FDTD simulations to ensure single-mode operation in the waveguides and ring resonator (inset of fig.\ref{fig:afm-elip}d) and the final film thickness and refractive index was determined using an ellipsometer.   

To define the structures, ARP-6200-09 electron beam positive resist was spin-coated and the patterns were formed using electron beam lithography. After exposure, the samples were developed and afterward etched using reactive ion etching (RIE Sentech Etchlab 200) with a mixture of SF$_6$ and O$_2$. A final layer of PMMA (>1 $\mu$m) or SiO$_{2}$ was used to enhance the confinement in the waveguides acting as a top cladding. To characterize the devices, we have used both edge and grating couplers. Fig.\ref{fig:afm-elip}d shows an electron microscope image of a cross-section of a device and fig.\ref{fig:afm-elip}e an optical microscope image of a device with grating couplers. For thermo-optic measurements, a thick (3 $\mu$m) layer of SiO$_{2}$ was deposited on top of the devices.

For the side  coupling configuration, we used a C-band tunable laser (Photonetics TUNICS-PRI 3642 HE 15). The polarization incident in the waveguide was selected using a free space polarizer and polarization-maintaining fibers (OZ Optics V-groove assembly). To obtain the transmission spectrum of the optical ring resonators, the wavelength of the laser was swept in the desired range with 1 pm resolution. The output power was recorded with a photodetector (Newport 843-R).

\section{RESULTS AND DISCUSSION}

\subsection{Device characterization}

Many ring resonators fabricated on PECVD and ICPCVD films with various parameters such as waveguide width, gap, ring radius, and deposition temperature were thoroughly studied and compared, and the overall results can be found in the supplementary information together with the equations to determine the quality factor and waveguide propagation losses.

The highest quality factors were obtained for a deposition with ICPVCD at a temperature of 150$^{\circ}$C and the data is shown in fig.\ref{fig:best_q}a, from which a free spectral range of 1.3 nm is determined. A loaded quality factor (Q$_L$) of 4.2x$10^{5}$ was measured and the intrinsic quality factor (Q$_{int}$) of the device is estimated to be 5.7x$10^{5}$ which is more than three times higher than previously reported results \cite{cmos-asic}, corresponding to waveguide propagation losses of 0.89 dB/cm. The lowest propagation loss was 0.78 dB/cm for the ring resonator shown in Fig.S6 of the supplementary information. 

\subsection{Thermo-optic coefficient of ICPCVD a-SiC}

\begin{figure*}
\includegraphics[width=1\textwidth]{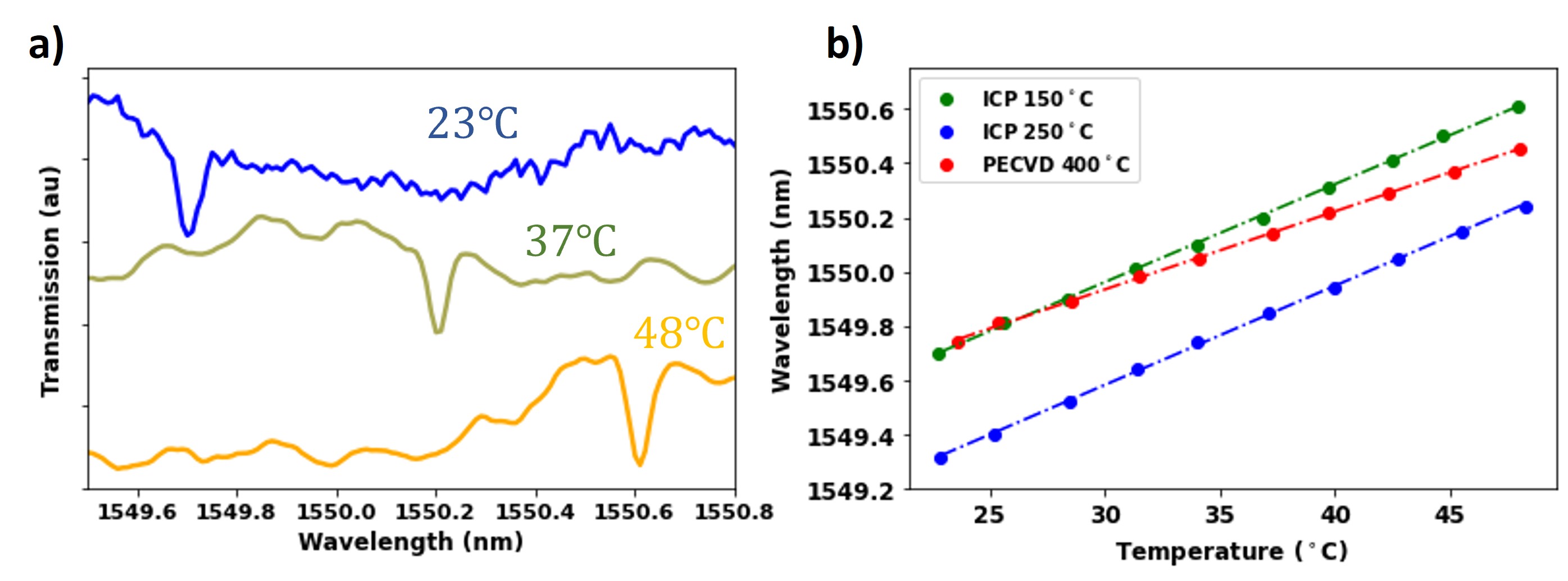}

\caption{\label{fig:thermo} a) Transmission spectra for a-SiC deposited at 150$^{\circ}$C as the temperature of the devices is raised at 23$^{\circ}$C (blue), 37$^{\circ}$C (green) and 48$^{\circ}$C (yellow) and b) wavelength shift of a resonance dip as a function of temperature for devices made on ICPCVD a-SiC films deposited at 150$^{\circ}$C and 250$^{\circ}$C and PECVD a-SiC deposited at 400$^{\circ}$C taken with steps of 2$^{\circ}$C.} 

\end{figure*}

The thermo-optic coefficient plays a major role in the choice of photonic platforms, where many applications require low-power thermal tuning to reduce the thermal cross-talk between devices. In this latter case, platforms based on SiN and SiO$_{2}$ have shown poor performance\cite{thermo-sio2}, making thermal tuning a challenging task. In this work, we measure the thermo-optic coefficient of a-SiC deposited via ICPCVD by studying the shift in the resonance wavelength of optical ring resonators upon a change in temperature in a range between 23$^{\circ}$C and 47$^{\circ}$C (the setup is shown in fig.S9 of the supplementary information). Fig.\ref{fig:thermo}a shows a representative transmission spectrum at different temperatures taken from the device from which we achieved the highest quality factors. In fig.\ref{fig:thermo}b is represented the value of a specific resonance dip for temperature steps of 2$^{\circ}$C for devices fabricated on ICP and PECVD films. The change in effective refractive index ($n_{eff}$) as a function of the material temperature can be derived from the following relation \cite{thermooptic}: 

\begin{equation}
\frac{d\lambda}{dT}=\left ( an_{\mathrm{eff}} + \frac{dn_{\mathrm{eff}}}{dT} \right )\frac{\lambda}{n_{g}}
\end{equation}

\noindent with a = 2.6x$10^{-6}/$$^{\circ}$C being the expansion coefficient of the thermal Silicon Dioxide upon a change in temperature, $n_{eff}$ the effective index of the a-SiC waveguide with 750 nm in width and varying thickness (measured by ellipsometry and confirmed with SEM) estimated using 3D FDTD simulations in Lumerical, and n$_{g}$ is the group index at 1550 nm that is obtained from the transmission spectra (see fig.S7 of the supplementary information). The equation that relates the thermo-optic coefficient of the materials involved with the change in the effective refractive index as a function of temperature was obtained in previous studies using the overlap integral approximation \cite{thermo1,thermo2}:

\begin{equation}
\frac{dn_{\mathrm{eff}}}{dT}=\Gamma_{\mathrm{SiO_{2}}}\frac{dn_{\mathrm{SiO_{2}}}}{dT} + \Gamma_{\mathrm{a-SiC}}\frac{dn_{\mathrm{a-SiC}}}{dT}
\end{equation}

\noindent where $\Gamma$ denotes the overlap integral coefficients for the Silicon Dioxide cladding and the Silicon Carbide waveguide and is determined using 3D Mode simulations in Lumerical with the specific dimensions of the individual devices and the thermo-optic coefficient of PECVD SiO$_{2}$ is already known to be 0.96x$10^{-6}/$$^{\circ}$C as determined in literature\cite{thermo-sio2}. For the devices made on a-SiC deposited at 150$^{\circ}$C, a thermo-optic coefficient of 7.3x10$^{-5}/^{\circ}C$ is obtained, which is three times higher than PECVD SiN \cite{thermo-sio2}.  As a reference, thermo-optic measurements of a-SiC deposited via PECVD at 400$^{\circ}C$ are also shown, with a thermo-optic coefficient of 5.1x10$^{-5}/^{\circ}C$, overall in agreement with previous works \cite{sirich-thermo}.

\begin{figure*}
\includegraphics[width=1\textwidth]{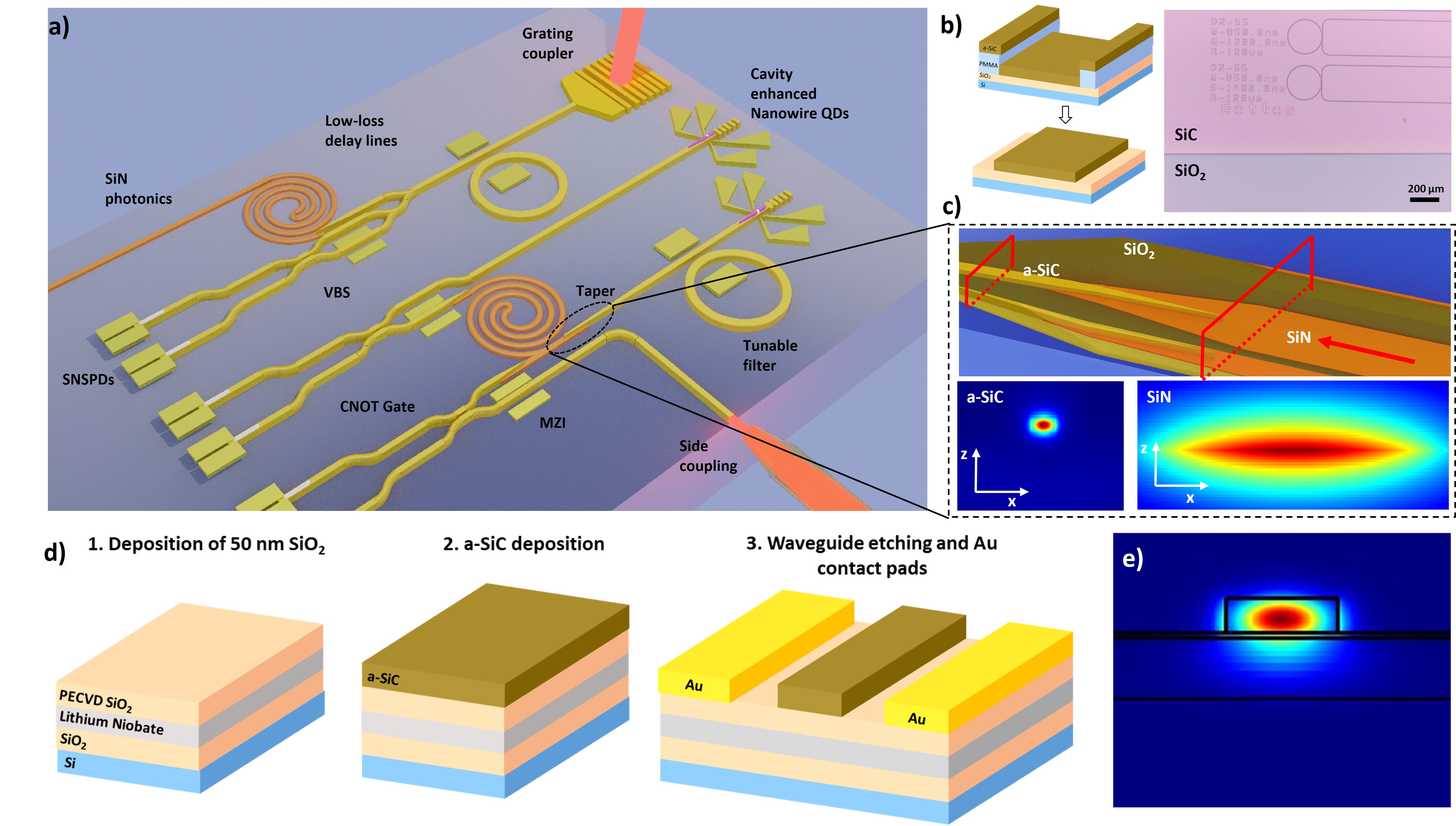}

\caption{\label{fig:asicln} a) Amorphous Silicon Carbide Photonic devices made on a Silicon Nitride platform. Light coupling can be performed via side coupling, using grating couplers or waveguide-embedded nanowire quantum dots. The representation includes tunable single-photon filtering with ring resonators, low-loss delay lines made on thin-film a-SiC (or SiN), Mach-Zehnder interferometers (MZI), Variable Beamsplitters (VBS) and Superconducting Nanowire Single Photon Detectors (SNSPDs). b) lift-off process and optical image of the fabricated devices, c) FDTD simulations of the mode profile in tapered a-SiC/SiN waveguides for high coupling efficiency, d) fabrication flow of a-SiC on LN for electro-optic modulation and e) mode profile obtained via FDTD 3D simulations of a-SiC/LN. The isolation layer between a-SiC and LN allows for etching of the a-SiC waveguides without affecting the performance of the LN}

\end{figure*}

\section{HYBRID INTEGRATION AND FUTURE OUTLOOK}

In quantum photonic circuits, routing photons with low losses is a vital requirement and to this end, over the last decade, material platforms such as LPCVD Silicon Nitride have been extensively optimized to reduce the losses. Recent works using thin film SiN waveguides have enabled high-yield and wafer-scale fabrication with losses as low as 1 dB/m\cite{thinfilm-sin1, thinfilm-sin2} which is a fundamental requirement for two-photon interference on chip \cite{HOM, Chanana2022}. As a future outlook, we highlight in fig.\ref{fig:asicln}a the use of 280 nm thick a-SiC in combination with low-loss waveguides based on Silicon Nitride thin-films (40 nm) and Lithium Niobate, two promising platforms for integrated quantum photonics. In this scheme, delay lines and photonic routing can be done with low-loss on SiN to later exploit the non-linearity of a-SiC in wavelength conversion experiments and generation of entangled photon pairs. It can also be used in combination with crystalline Silicon Carbide or Silicon to deterministically address single photon sources and route the single photons. The input light can be delivered to the photonic structures using grating couplers, edge couplers or directly produced on-chip by embedded nanowire quantum dots based on III-V materials\cite{deterministic}. A tapered waveguide is designed to avoid coupling losses when the light is injected from 40 nm SiN waveguides to the 280 nm a-SiC. These structures are later covered with a 3$\mu$m thick silicon dioxide cladding that is also tapered to improve the confinement. From FDTD simulations, a tapper length of 150 $\mu$m achieves a coupling efficiency of 92.6$\%$ at a wavelength of 1550 nm with very high confinement in the a-SiC waveguide confirmed by the low bending losses of the mode (see fig.S12 in the supplementary information). Manipulation of light can be performed via variable beamsplitters and Mach-Zehnder interferometers and routed towards superconducting single-photon detectors forming the basic building blocks of a photonic CNOT gate.

Owing to the low temperature at which these films are deposited, we demonstrate a feasible approach to integrate the a-SiC films with current platforms based in a lift-off process with PMMA in fig.\ref{fig:asicln}b, where the specific details about the procedure can be found in the supplementary information. 

Lithium Niobate (LiNbO$_3$) provides efficient electro-optic modulation, high second-order non-linearity, broad transparency window from the visible to the mid-infrared range \cite{lithium-niobate} and ultralow losses at telecommunication wavelengths as demonstrated in an LNOI platform \cite{thinfilm-LN}. For this reason, to include our a-SiC devices in this platform and combine the properties that they both offer, in fig.\ref{fig:asicln}d we demonstrate a fabrication route for heat-free tuning of photonic devices together with FDTD simulations of the mode profile (fig.\ref{fig:asicln}e), where the oxide isolation layer protects the Lithium Niobate from the etching of a-SiC waveguides. A more detailed overview of the process is shown in the supplementary information. 

\section{Conclusions}

Amorphous Silicon Carbide photonic devices with very low losses have been demonstrated with intrinsic quality factor reaching 5.7x$10^{5}$, corresponding to waveguide propagation losses between 0.78-1.06 dB/cm. The film deposition was optimized at a temperature of 150$^{\circ}$C for heterogeneous integration for a variety of photonic platforms without affecting the overall optical properties. In addition, we characterized the thermo-optic coefficient of the fabricated devices, with a TOC for ICPCVD a-SiC of 7.3x10$^{-5}/^{\circ}C$ providing better performance than similar devices fabricated on a PECVD SiN platform. 
 
Further investigation is needed to understand the fundamental properties of the films and improve their quality. Besides geometrical losses, such as bending losses and sidewall roughness, there could be other contributions such as the density of the films, refractive index changes, stress in the films, hydrogen bonds with silicon and carbon and microcrystallization of Si, C or SiC forming scattering clusters.

The low-temperature deposition-enabled photonic integration using a simple lift-off process can be implemented without damaging other components in the chip. By designing a tapered interface between SiN and a-SiC we obtain an optical coupling of 92.6$\%$. Exciting opportunities are possible by combining ICPCVD a-SiC films with CMOS compatible processes, hybrid photonic structures involving III-V materials, and current photonic platforms such as SiN for low-loss routing of photons or LiNbO$_3$ for heat-free electro-optical modulation. 

\section{Backmatter}

\begin{backmatter}

\bmsection{Acknowledgments}
R.K, D.P. and B.L.R. optimized the deposition recipe and characterized the films. B.L.R. fabricated the devices and coordinated the work. B.L.R., S.A. and N.S. performed optical measurements of the devices and analyzed the data. B.L.R. and Z.L. performed Lumerical simulations. T.S. helped in the assembly and adaptation of the optical setups. All authors contributed equally to the manuscript. I.E.Z., S.F.P. and H.B. conceptualized the experiment. J.C. and S.G. assisted in optical measurements. N.S. acknowledges the NWO OTP COMB-O project (18757).

\bmsection{Disclosures}
The authors declare no conflicts of interest.

\bmsection{Data availability} The data and fabricated samples supporting this study are available from the corresponding authors for further analysis upon reasonable request.  

\bmsection{Supplemental document}
See Supplementary information for supporting content. 

\end{backmatter}


\bibliography{Optica-template}

\end{document}